\documentclass{article}

\usepackage[numbers]{natbib}
\usepackage{arxiv}

\usepackage[utf8]{inputenc} %
\usepackage[T1]{fontenc}    %
\usepackage{hyperref}       %
\usepackage{url}            %
\usepackage{booktabs}       %
\usepackage{amsfonts}       %
\usepackage{nicefrac}       %
\usepackage{microtype}      %
\usepackage{lipsum}		%
\usepackage{graphicx}
\usepackage{doi}
\usepackage{multirow}
\usepackage{cleveref}
\usepackage{array}
\usepackage[normalem]{ulem}

\title{Toward Responsible AI Use: \\Considerations for Sustainability Impact Assessment}

\author{{Eva Thelisson}\thanks{Corresponding author} \\
    AI Transparency Institute \\
    Switzerland \\
    \texttt{eva@aitransparencyinstitute.com} \\
	\And
	{Grzegorz Mika} \\
	Telecom SudParis\\
	Institute Polytechnique de Paris \\
    France\\
    \texttt{grzegorz.mika@ip-paris.fr} \\
	\And
	Quentin Schneiter \\
	AI Transparency Institute \\
	Switzerland \\
	\AND
	Kirtan Padh \\
	Helmholtz Munich, Helmholz AI \\
    MCML, ELLIS, TU Munich \\
    Germany \\
    \texttt{kirtan.padh@tum.de} \\
    \And
	Himanshu Verma \\
	TU Delft \\
	Netherlands \\
	\texttt{H.Verma@tudelft.nl} \\
}

\renewcommand{\shorttitle}{Toward Responsible AI Use}

\hypersetup{
pdftitle={Toward Responsible AI Use: Considerations for Sustainability Impact Assessment},
pdfsubject={q-bio.NC, q-bio.QM},
pdfauthor={David S.~Hippocampus, Elias D.~Striatum},
pdfkeywords={Sustainability, Green AI, ESG Digital and Green, Environmental Impact, Societal Impact, Governance},
}

\begin{document}
\maketitle

\begin{abstract}
	As AI/ML models, including Large Language Models, continue to scale with massive datasets, so does their consumption of undeniably limited natural resources, and impact on society. In this collaboration between AI, Sustainability, HCI and legal researchers, we aim to enable a transition to sustainable AI development  by enabling stakeholders across the AI value chain to assess and quantitfy the environmental and societal impact of AI.  We present the ESG Digital and Green Index (DGI), which offers a dashboard for assessing a company's performance in achieving sustainability targets. This includes monitoring the efficiency and sustainable use of limited natural resources related to AI technologies (water, electricity, etc). It also addresses the societal and governance challenges related to AI. The DGI creates incentives for companies to align their pathway with the Sustainable Development Goals (SDGs). The value, challenges and limitations of our methodology and findings are discussed in the paper.
\end{abstract}


\section{Introduction}

Algorithmic systems are becoming pervasive in daily life, significantly influencing decision-making~\cite{stuart2019human} and service delivery across the public~\citep{medaglia2023artificial} and private sectors\cite{addad2020face,beck1996applications,shaheen2021applications,faught1986applications, zhang2021artificial}. Large Language Models (LLMs), such as ChatGPT\cite{kasirzadeh2023chatgpt}, exemplify this trend, but their deployment raises substantial environmental~\citep{hoang2019fabuleux} and ethical~\citep{bender2021dangers} concerns. These systems demand hefty computational power, consequently escalating energy consumption and greenhouse gas emissions\cite{luccioni2023counting}. Moreover, their responsible and ethical development is crucial to avoid perpetuating biases or disseminating misinformation\cite{floridi2018ethical,wang2022against,buolamwini2018gender}. As machines take on more responsibility, they must be designed to be more ethically responsible~\citep{wallach2010moral} and socially and robustly beneficial~\citep{hoang2019fabuleux}. 

In response to these challenges, regulatory structures are emerging globally. For instance, the sector-specific governance framework developed by the EU Commission (EU AI Act), UNESCO with its Recommendation on the Ethics of AI~\cite{unesco2021recommendation} and the OECD via specific Principles on AI. The EU has also implemented sustainability reporting standards under Directive 2013/34/EU (amended by Directive 2022/2464), obliging businesses to disclose information pertaining to their sustainability efforts and environmental impacts, promoting a clearer understanding of how such issues influence business evolution. On 31 July 2023, the European Commission (EC) adopted the delegated act of the European Sustainability Reporting Standards (ESRS), under which companies must publish information about their impact on the environment as well as the impact of environmental and social issues on their financial risks and opportunities.

\begin{figure}[!t]
    \centering
    \includegraphics[width=0.75\textwidth]{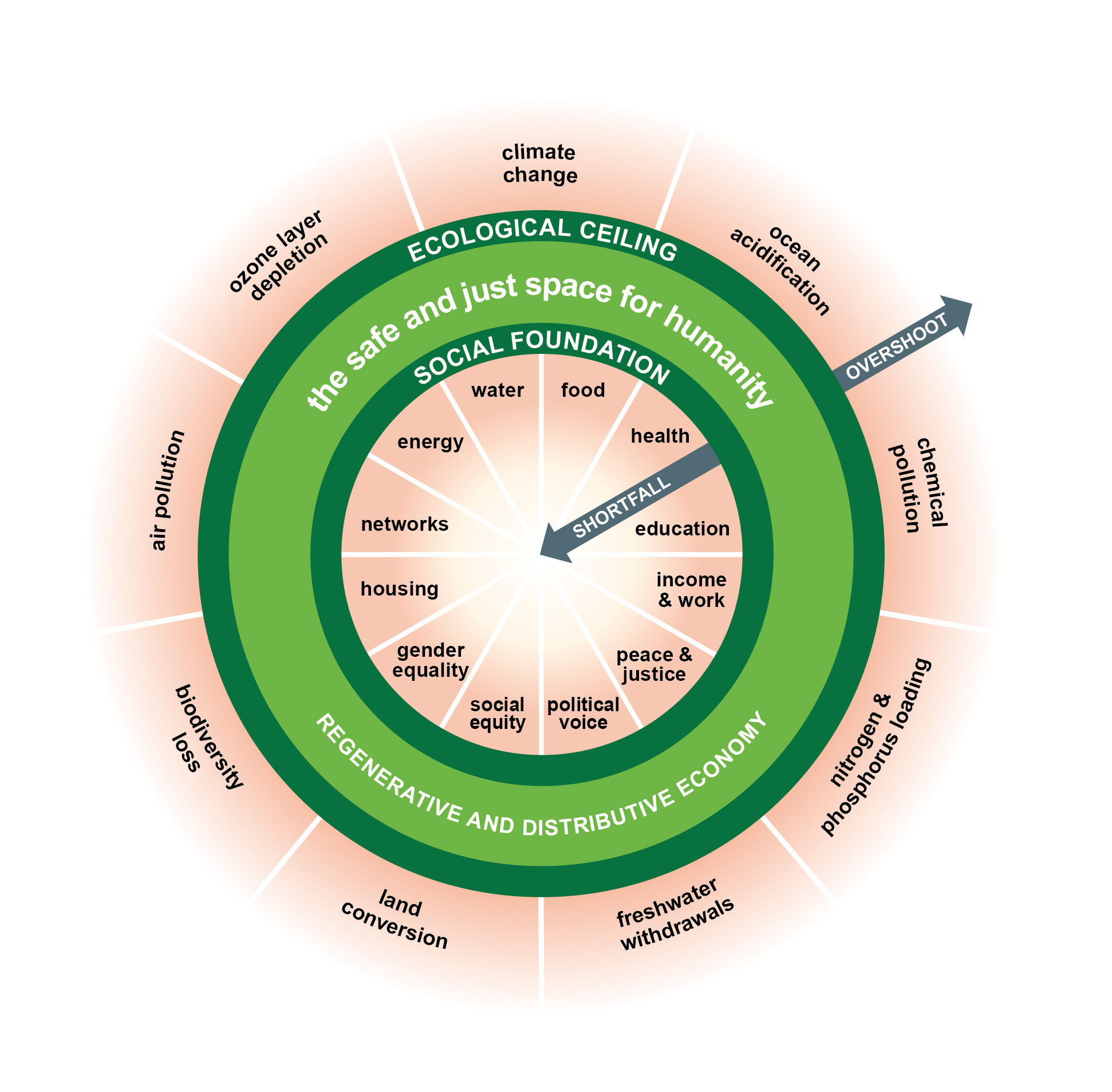}
    \caption{Doughnut Theory (By DoughnutEconomics - Own work, CC BY-SA 4.0, \url{https://commons.wikimedia.org/w/index.php?curid=75695171})}
    \label{fig:donut_economics}
\end{figure}

Addressing the urgent regulatory and societal need for diligent sustainability impact monitoring, we present the ESG Digital and Green Index (DGI), a novel analytical instrument designed to aid organizations in aligning their technological advancements with global environmental sustainability objectives, including adherence to the Sustainable Development Goals (SDGs) and the Paris Climate Agreement~\cite{glanemann2020paris}. The DGI is grounded in Kate Raworth's Doughnut model~\cite{raworth2017doughnut}. This visual framework for economic sustainability (see \cref{fig:donut_economics}) presented as a doughnut combines the concept of planetary limits with the complementary concept of social boundaries. The model proposes to consider the performance of an economy by the extent to which people's needs are met without exceeding the Earth's ecological ceiling, an economic compass to guide the economy and our societies. This paper refers to this compass for digital ecosystems. The DGI offers a dashboard for evaluating company performances regarding sustainable AI development and is the first metric for AI's impact on limited resources. 

The DGI was prepared following a multi-disciplinary co-design. It went through a series of revisions to improve the initial conceptual and methodological framework through multiple iterative steps. This step-wise scientific approach and consultative process involved participants with expertise in design-thinking, Human-Computer Interactions (HCI), sustainability, data processing, AI, and law, capturing the complex and multi-dimensional relationship between AI and sustainable development. The DGI comprises three dimensions: environmental ceiling, social floor, and transverse. The indicators for environmental ceilings include climate change, the use of natural resources, pollution, and disruption of natural cycles, biodiversity, and ecosystems. The indicators for the social floor include governance, health and well-being, education, and security. The indicators for transverse include global sustainability.

More specifically, we aim to explore the following research questions: \textit{How can we bring about a silent transformation of business models to align the interests of all stakeholders? In other words, from the limitless growth of digital technologies to a balanced prosperity that takes into account the limits of the Earth's natural resources and the internal limits of human beings and human rights?}

While answering the research question, we make the following key contributions:
\begin{itemize}
    \item We raise up the level of awareness on the possible use of the Doughnut's compass for AI and digital activities. 
    \item We offer a positive economic perspective, able to return to a prosperous equilibrium.
    \item Finally, we offer an online application to be tested by companies and organisations.
\end{itemize}

The remainder of this paper is organized as follows. \cref{sec:literature} provides an overview of the current discourse on AI and creates a link with the challenges of sustainable growth and societal impact and how this relates to the development of the DGI. Then, \cref{sec:dgi} presents our contribution, namely, the ESG Digital and Green Index (DGI). Finally, we discuss the value and shortcomings of applying the DGI framework in light of current sustainability challenges in \cref{sec:discussion}.

\section{Literature Review}
\label{sec:literature}
In this section, we first review the literature on AI and its carbon footprint and societal impact (\cref{sec:background-AI}), followed by a review of the literature on sustainability and the context in which the DGI fits into this discourse (\cref{sec:background-sustainability}), and finally related regulations on green AI (\cref{sec:background:green-deal}).

\subsection{Artificial Intelligence}
\label{sec:background-AI}

Artificial intelligence (AI) is characterised by an heterogeneity of technologies, usages and access to different age groups resulting in different types of interactions between the different types of users (children, teenagers, adults, elderly people). 

This heterogeneity offers a wide space of characteristics which are difficult to represent as a whole and made standardisation processes difficult. In this context, this is an important factor to take into consideration.

\begin{figure*}[!t]
    \centering
    \includegraphics[width=.6\textwidth]{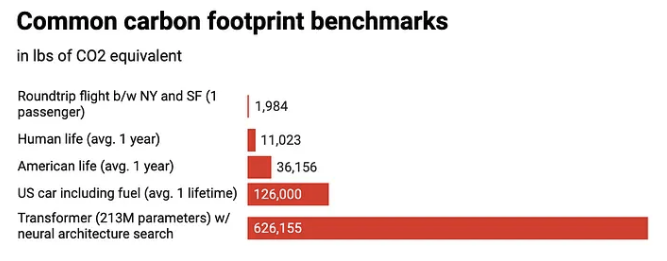}
    \caption{CO2 carbon footprint benchmarks. Chart MIT Technology Review and source~\cite{strubell2019energy}.}
    \label{fig:carbon_footprint}
\end{figure*}

Significant advancements have been driven by methods that harness computational power to analyze extensive datasets and generate patterns and insights, as indicated by Rich Sutton in "The Bitter Lesson"~\cite{sutton2019bitter}. Moreover, AI techniques are changing our lives and transforming every sector of the economy \cite{sejnowski2018deep, jordan2015machine} and impacting on society \cite{amodei2016concrete}. However, the surge in computational capacities presents considerable challenges, especially in the context of environmental sustainability~\cite{verdecchia2023systematic, utzclimate}. Generative AI models, particularly large language models, have been expanding rapidly, exemplified by the forthcoming 2024 frontier expected to exceed 100 trillion tokens, ushering in heightened computational demands and carbon footprints~\cite{zhuang2023survey}. \cref{fig:carbon_footprint} delineates the substantial CO2 emissions associated with advanced AI models compared to daily human activities.

Furthermore, generative and general-purpose AI poses a host of new challenges, including Nonconsensual intimate imagery (NCII)~\citep{kobriger2021out}, which can have potentially traumatic and life-changing effects~\cite{hao2021deepfake}. Voice or video cloning scams using generative AI are also on the rise~\citep{stupp2019fraudsters, verma2023they}, as is disinformation~\citep{goldstein2023generative, diresta2020supply}. The seminal paper on the risks of LLMs by Bender, Gebru et al. ~\citep{bender2021dangers}, written in 2020, already talked about ``weighing the environmental and financial costs first, investing resources into curating and carefully documenting datasets rather than ingesting everything on the web, carrying out pre-development exercises evaluating how
the planned approach fits into research and development goals and
supports stakeholder values''.

Addressing these concerns, this paper introduces a novel research field termed `computational sustainability,' urging a reevaluation of AI advancements' environmental and social repercussions, also taking into account reports from technology giants \cite{Google2023Environmental, Microsoft2023Environmental}.

\subsection{Sustainability}
\label{sec:background-sustainability}

\begin{figure*}[!ht]
    \centering
    \includegraphics[width=.75\textwidth]{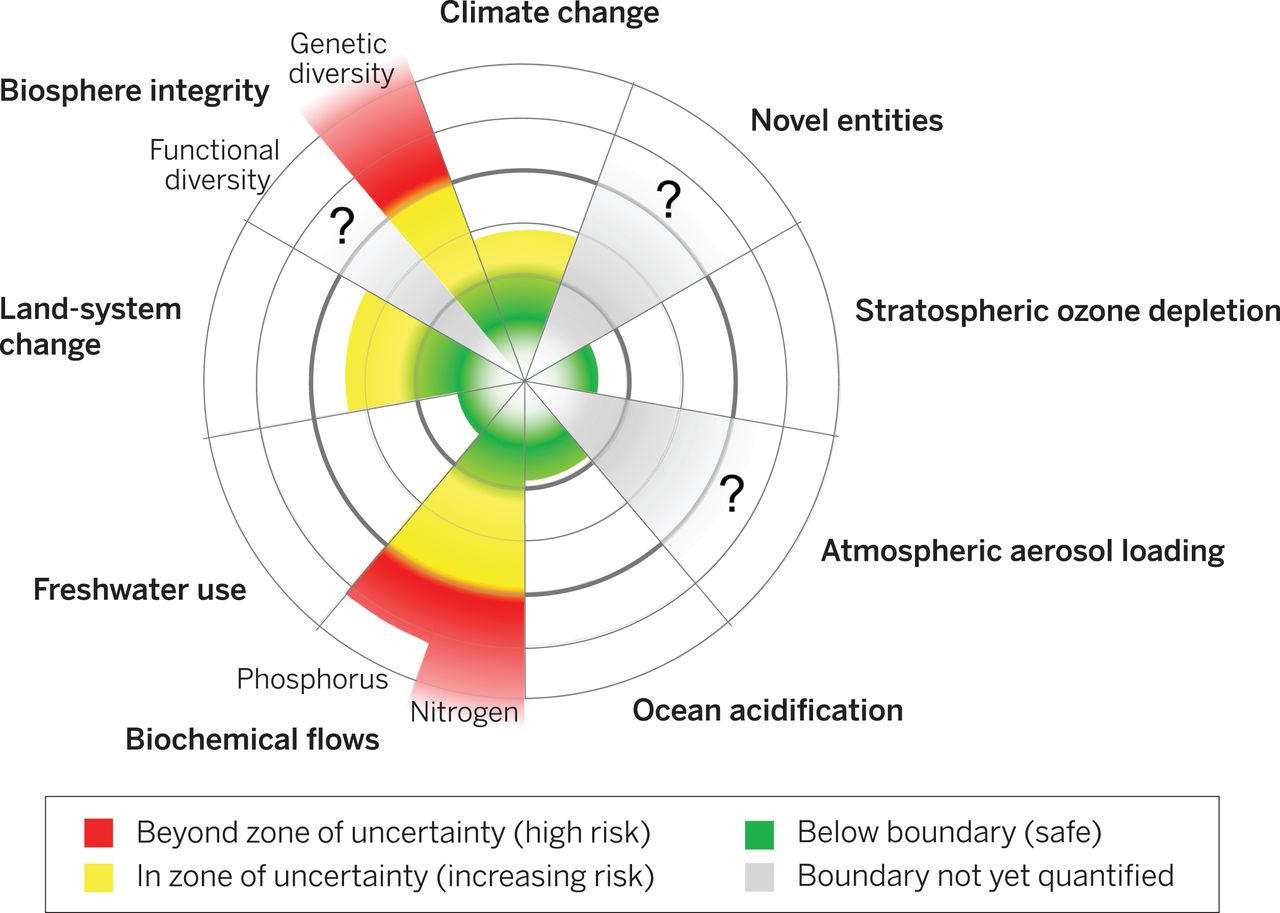}
    \caption{Planet boundaries~\citep{rockstrom2009planetary}}
    \label{fig:planet_boundaries}
\end{figure*}

Historically, sustainability refers to the concept of sustainable development introduced by the Brundtland Report~\citep{brundtland1987our} as a development that meets the needs of the present without compromising the ability of future generations to meet their own needs. While this definition still garners international consensus, it has since been complemented by various frameworks, including the 17 SDGs adopted by the 193 United Nations member states on September 25, 2015~\citep{nations2015transforming}.
The SDGs serve as a model to consider sustainability in its entirety, addressing environmental, social, and economic aspects. However, these goals, which refer to so-called weak sustainability that admits the fungibility of natural, human, and economic capitals, benefit from being complemented by models that specify our environment's physical thresholds while prioritizing humans and their needs.
To obtain a framework consistent with the physical realities characterizing the impact of human societies on the environment, it is interesting to integrate the concept of planetary boundaries and its interpretation by the economist Kate Raworth~\cite{raworth2017doughnut}.

The nine planetary boundaries (\cref{fig:planet_boundaries}) relate to the study by~\citet{rockstrom2009planetary}, initially published in 2009, reissued in 2015, which specifies both the critical areas to consider for maintaining the stability of the Earth system and their overshoot thresholds and current impact levels. The resulting model, widely cited within the scientific community, provides a solid foundation for distinguishing the impact categories to monitor and their criticality for a sustainability assessment.

The planetary boundaries are incorporated into the Doughnut theory (\cref{fig:donut_economics}), developed by the economist Kate Raworth for her eponymous book published in 2017~\citep{raworth2017doughnut}. The model embeds the thresholds in a simplified economic model depicting the "safe and just space for humanity" within which our systems' development is constrained. The model consists of a ``social foundation'', subdivided into criteria ensuring minimum living conditions for humans, a ``governance ceiling'' and an ``environmental ceiling'' set by the thresholds of the planetary boundaries.

Based on the Doughnut model, the DGI aligns with so-called strong sustainability, where ecological limits delineate the social space, and both jointly define the space in which economic activities can operate sustainably. This sustainability goal is formally recognised in the Article 191 of the Treaty on the Functioning of the European Union (TFEU). The Article 191 is not a sectorial-specific clause. It applies in any sector, including the ones using AI-based technologies.

\subsection{European Green Deal}
\label{sec:background:green-deal}
The purpose of the European Green Deal is for the European Union to become the world's first “climate-neutral bloc” by 2050. It has goals extending to many different sectors. The plan includes potential carbon tariffs for countries that don't curtail their greenhouse gas pollution at the same rate. It also includes a circular economy action plan. The European Commission has released many publications on circular economy, including one that requires Member States to carry out activities related to changing their economies into circular economies. The Circular Economy has indeed become a vital component of the European Green Deal. The goal is to create a sustainable, low-carbon, resource-efficient, and competitive economy in the Single Market.

The EU plans to finance the policies set out in the Green Deal through an investment plan, which forecasts at least €1 trillion in investment. 

\subsection{EU Regulation on Corporate Due Diligence}
\label{sec:eu-regulation}

The European Parliament recommended a corporate due diligence and accountability proposal in March 2021. The Parliament supports mandatory legislation due to the limited progress of voluntary standards. The proposed directive requires companies to conduct due diligence for human rights, environment, and governance impacts. Therefore, we build our tool around these components. Under EU Law, sanctions and liability regimes will be enforced for non-compliance.

The Corporate Sustainability Due Diligence Directive introduces duties for the directors of the EU companies covered. These duties include setting up and overseeing the implementation of the due diligence processes and integrating due diligence into the corporate strategy. In addition, when fulfilling their duty to act in the company's best interest, directors must consider the human rights, climate change, and environmental consequences of their decisions.

This Directive will apply to large EU limited liability companies with more than 500 employees and a net EUR 150 million exceeding turnover worldwide (Group 1). It will also impact a second group of companies in high-impact sectors, having more than 250 employees exceeding a net EUR 40 million turnover worldwide and operating in defined high-impact sectors (Group 2).
 
This regulation will also have an extraterritorial effect and apply to non–EU companies of both groups. It will also apply to global south companies active in the EU with turnover threshold aligned with Group 1 and 2, generated in the EU.  
The rules on corporate sustainability due diligence will be enforced through administrative supervision: Member States will designate an authority to supervise and impose effective, proportionate, and dissuasive sanctions, including fines and compliance orders. At the European level, the Commission will set up a European Network of Supervisory Authorities that will bring together representatives of the national bodies to ensure a coordinated approach. Regarding Civil liability, Member States will ensure that victims get compensation for damages resulting from failing to comply with the obligations of the new proposals. The rules of directors' duties are enforced through existing Member States' laws. 

Private Enforcement mechanisms of the Corporate Sustainability Due Diligence Proposal Directive may reinforce adequate protection of the environment and human rights. This legislation gives rise to the need for a level playing field for companies to avoid fragmentation and provide legal certainty for businesses operating in a single market. A level playing field requires information, transparency, and high compliance with the obligation to disclose sustainability measures under the Sustainability Due Diligence Directive.

The ESG Digital and Green Index (DGI) is embedding these obligations. This novel analytical instrument is crafted to assist both the public and private sectors in discerning and adopting environmentally viable trajectories in the digital era. This initiative is primed to facilitate the realization of global sustainability objectives, encompassing those delineated in the SDGs.

It includes all aspects of Annexes 1 and 2 of the EU Regulation on Corporate due Diligence on Environmental Aspects, Social Responsibility, and Governance Practices (Environmental Reporting, Social Responsibility Reporting, and Governance Reporting)~\cite{2022:eu:regulation1,2022:eu:regulation2}. It also includes Glossary Definitions as requested in Annex 2 of EU Regulation. All aspects of these Annexes are part of our ESG Digital and Green Index.

\section{The design of ESG Digital and Green Index}
\label{sec:dgi}

\begin{figure}[!tb]
  \centering
\includegraphics[width=0.8\linewidth]{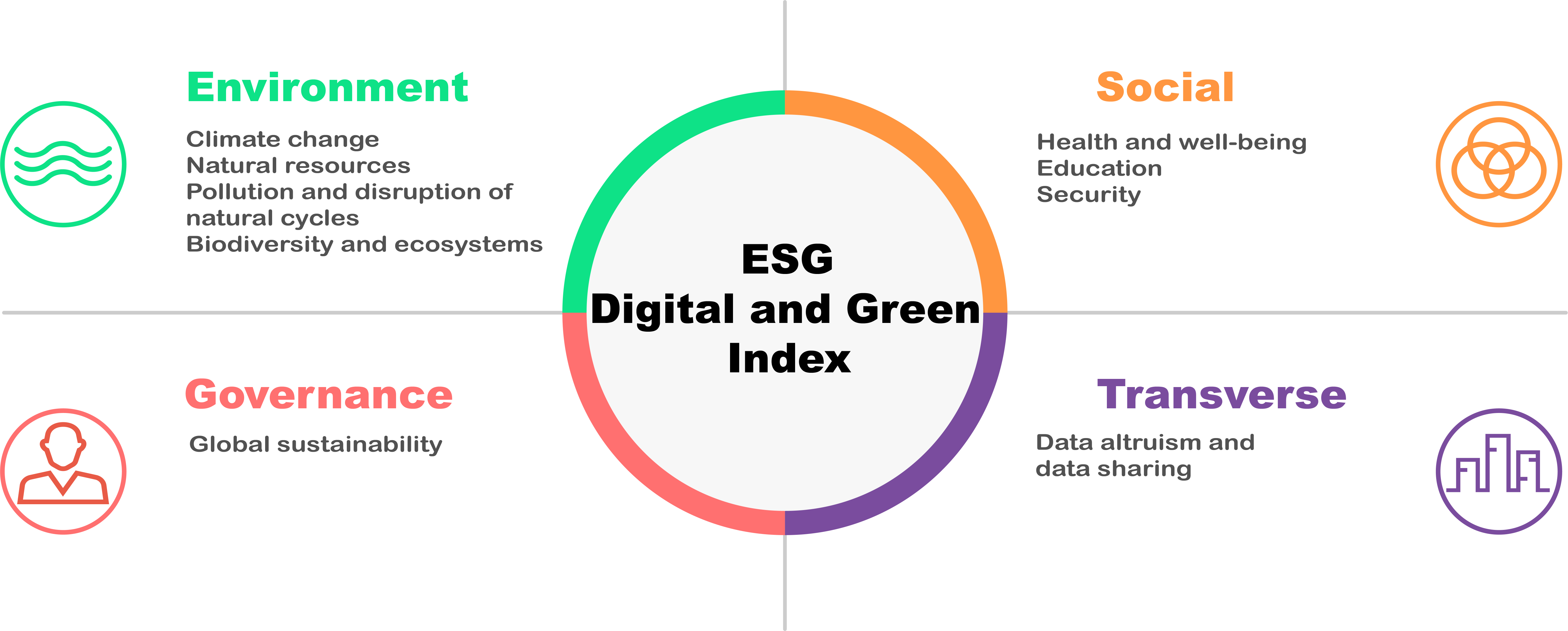}
  \caption{Conceptual framework for the ESG Digital and Green Index}
  \label{fig:framework}
\end{figure}

The ESG Digital and Green Index (DGI) is a development approach that seeks to deliver AI services and products that are both environmentally sustainable and socially beneficial. It offers a comparative advantage to AI-based tech services and products that are low carbon and climate resilient, mitigate, prevent, or even contribute to remediate pollution and maintain healthy and productive ecosystems. Toward the goal of sustainable artificial intelligence, we developed a comprehensive set of indicators for assessing the DGI. This section describes the development process and highlights the insights we gained. This definition emphasizes four closely interlinked concepts: low-carbon economy, health ecosystem, resilient society, and inclusive growth. 

\subsection{Design Rationale}

Through the DGI, we aim to provide a composite index to measure, track, and communicate digital and green performance. The DGI can raise awareness and sustain digital and green AI momentum in the public and private sectors. Because the index is based on a robust sustainability framework, it can highlight the SDG achievements linked to digital and green AI. Moreover, our index improves current knowledge of green AI and its drivers. The DGI provides an interactive learning experience and enhances users’ knowledge of green AI and strategy development. Because the tool can be used to simulate and understand the impacts of different policy and investment options on green AI performance, it can provide input in planning and supporting the formulation of green AI policies in key sectors.

\begin{table*}
\footnotesize
  \caption{Definitions of the indicator categories}
  \label{tab:commands-defs}
  \begin{center}
  \begin{tabular}{ m{15em} | m{8cm} }
    \toprule
    Indicator categories & Definitions\\
    \midrule
    Climate change & {Refers to the management and mitigation of greenhouse gas emissions from the activities of the evaluated organization. Within the context of AI utilization, the primary emphasis is on greenhouse gas emissions associated with energy consumption. This encompasses both scope 1 and scope 2 emissions, integrating emissions produced by the primary energy source for the generation of the consumed electricity} \\
    \midrule
    Natural Resources & {Refers to the management and reduction of non-renewable natural resource consumption by the evaluated organization's activities. The primary focus is on the tangible components constituting the infrastructure supporting AI, both within and external to the evaluated organization.}\\
    \midrule
    Pollution and disruption of natural cycles & {Refers to the management and limitation of synthetic substance releases from the evaluated organization's activities to preserve natural cycles.}
    \\
    \midrule
    Biodiversity and Ecosystems & {Refers to the management and limitation of impacts from the evaluated organization's activities that may lead to biodiversity loss, ecosystem degradation, and the loss of natural habitats.}\\ 
    \midrule
    Health and Well-being & {Refers to the management of employee well-being within the evaluated organization, especially considering the novel challenges posed by AI utilization.} \\
    \midrule
    Education & {Refers to continuous training and awareness-raising on aspects concerning sustainability and AI within the evaluated organization.}\\
    \midrule
    Security & {Refers to the management of all security aspects surrounding AI utilization within the evaluated organization.}\\
    \midrule
    Governance & {Refers to the implementation of collaborative, fair, accountable, and transparent governance within the evaluated organization, particularly in relation to the management of AI and the sustainability framework.} \\
    \midrule
    Global sustainability & {Refers to the cross-cutting sustainability monitoring, strategies, and actions within the evaluated organization.} \\
    \bottomrule
  \end{tabular}
  \end{center}
\end{table*}

\begin{figure}[h]
  \centering
  \includegraphics[width=0.7\linewidth]{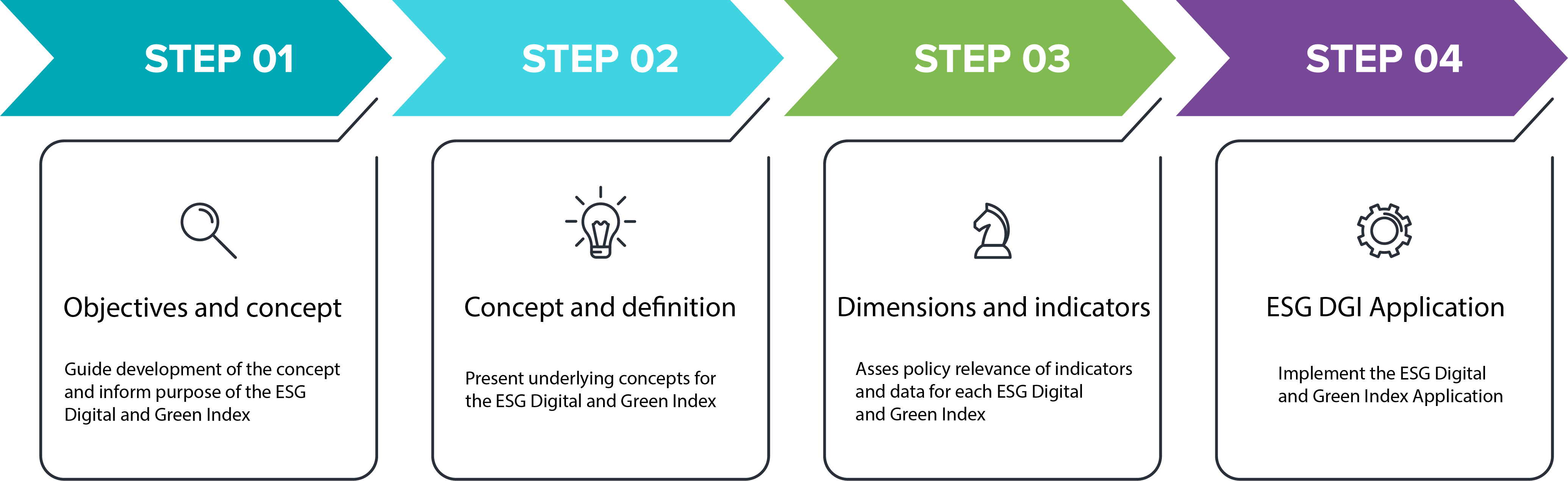}
  \caption{Stepwise approach to developing the ESG Digital and Green Index (phase I).}
  \label{fig:steps}
\end{figure}

\subsubsection{Global sustainability targets}

Our engagement to support the transformation of a company's economy cuts across different development issues. However, to maximize the impact of its products and services, we emphasize change in four priority areas: sustainable energy and water, responsible governance of sustainability and AI, and the health and well-being of employees. SDGs are an excellent framework for the transition towards a responsible innovation growth pathway for the public and private sectors. The DGI offers metrics to support a virtuous cycle creation for growth performance, which benefits multi-stakeholders. SDG indicators are a reliable and comprehensive dataset that provides an excellent source for constructing the DGI. 

Moreover, the DGI uses the foundations of the SDGs to build a new assessment mechanism to evaluate the impact of AI under several dimensions. To sum up, the criteria for selecting the sustainability targets are based on the SDG indicators and targets, both explicit and implicit, which were suggested in the OECD  report were used \cite{oecd2008handbook}. 

\subsection{System Description}

\begin{figure}[!bh]
  \centering
  \includegraphics[width=0.9\linewidth]{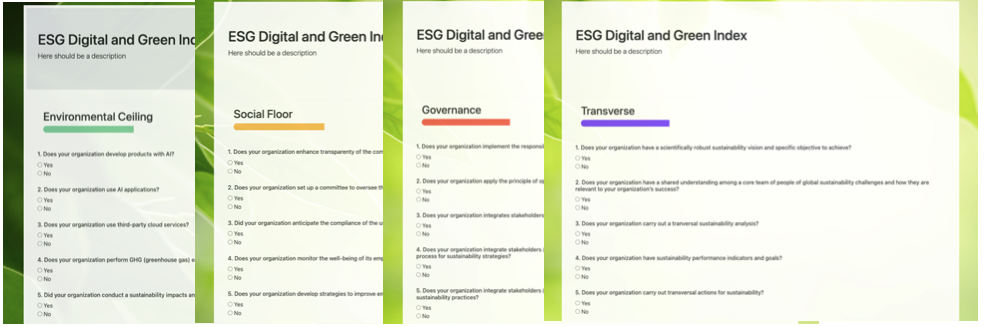}
  \caption{Example screenshot of the ESG Digital and Green Index application.}
  \label{fig:application}
\end{figure}

The design processes focus on steps to develop and apply the framework, such as an index and dashboards, and the range of institutions included in the development process. There are two general processes for designing green AI growth conceptual frameworks based on the fit-for-purpose principle and stakeholder consultations. The OECD Handbook~\cite{oecd2008handbook} suggests adopting a fit-for-purpose principle when selecting indicators that target end users’ needs. Because it entails a process that is entirely internal to organizations, developing the framework depends on a solid theoretical foundation, a well-defined narrative, and a scientifically driven set of indicators.

According to \cite{greco2019methodological}, composite indices involve a long sequence of steps that must be followed meticulously. We applied a stepwise approach to enhance the transparency, replicability, and credibility of the DGI. The first phase of our initiative consists of four steps illustrated in \cref{fig:steps}. Phase I consists of identifying and applying models which provide interlinkages among the indicators and require data, presents underlying concept and assess indicator weights. Then, the last step includes the DGI app development to calculate the green scoring for use in Phase II by collecting feedback. The index scoring is based on the responses provided in the survey. We have created an appropriate, refined database of questions for each indicator, e.g., climate change includes sample questions: \textit{"Does your organization develop products with AI?"}, if an answer would be positive, then \textit{"Does your organization develop products with AI?"} (Fig \ref{fig:application}).

The DGI application builds on the four dimensions of sustainable AI – environmental ceiling, social floor, governance, and transverse (\cref{fig:framework}). These dimensions are closely interlinked based on the concepts of low carbon economy, resilient society, ecosystem health, and inclusive growth. Each dimension in the DGI contains the defined indicator categories (\cref{tab:commands-defs}). These indicator categories are essential to transitioning to green and sustainable AI pathways. The environmental ceiling covers climate change (e.g., carbon emissions and their reduction strategy and actions), natural resources (e.g., resource consumption, electronic waste, and repair policy), pollution (e.g., electronic waste management policy and monitoring), and biodiversity (e.g., biodiversity impact and ecosystems). The social floor dimension includes health and well-being (e.g., well-being monitoring, adaptation to change), education (e.g., stakeholder education, sustainability issues and practices), and security (e.g., privacy, risk assessment). The transverse includes global sustainability based on vision, company culture, life cycle, and favorable impact products reducing carbon emission. The \cref{tab:commands-framework} shows the entire index in more detail.

\begin{table*}[ht!]
\footnotesize
  \caption{Indicator framework for the ESG Digital and Green Index}
  \label{tab:commands-framework}
  \resizebox{15cm}{!}{
  \begin{tabular}{c | c | l }
    \toprule
    Dimensions & Indicator categories & Indicators\\
    \midrule
    \multirow{24}{*}{Environmental ceiling} & \multirow{6}{*}{Climate change} & Carbon emissions associated with the development the AI software\\
    & &Carbon emissions associated with the usage of AI software (inside the audited company)\\
    & &Carbon emissions associated with the usage of AI software (outside the audited company)\\
    & &Carbon accounting\\
    & &Carbon reduction strategy\\
    & &Carbon reduction actions\\ \cline{2-3}
    & \multirow{10}{*}{Natural resources} & Resource consumption associated with the development of the AI software\\
    & &Resource consumption associated with the usage of the AI software (inside the audited company)\\
    & &Resource consumption associated with the usage of the AI software (outside the audited company)\\
    & &Electronic waste recycling policy\\
    & &Electronic repair policy\\
    & &Computer equipement sharing policy (pooling)\\
    & &Computer equipment external reusage\\
    & &Natural ressources usage accounting\\
    & &Strategy to reduce natural ressources usage\\
    & &Actions to reduce natural ressources usage\\ \cline{2-3}
    & \multirow{5}{*}{Pollution and disruption of natural cycles} &Consideration of pollution in the sustainability strategy\\
    & &Pollution accounting\\
    & &Pollution reduction actions\\
    & &Electronic waste management policy\\
    & &Monitoring of electronic waste\\ \cline{2-3}
    & \multirow{3}{*}{Biodiversity and ecosystems} &Consideration of biodiversity in the sustainability strategy\\
    & &Biodiversity impact accounting\\
    & &Biodiversity impact reduction actions\\
    \midrule
    
    \multirow{13}{*}{Social floor} & \multirow{5}{*}{Health and well-being} & Well-being monitoring of the employees\\
    & &Well-being strategy\\ 
    & &Well-being actions\\
    & &Impact assessments on effect of AI systems\\
    & &Adaptation to change\\ \cline{2-3}
    & \multirow{4}{*}{Education} &Stakeholders education about sustainability issues\\
    & &Stakeholders education about sustainability pratices\\
    & &Stakeholders education about AI\\
    & &Stakeholders training to use AI\\ \cline{2-3}
    & \multirow{4}{*}{Security} &Data security\\
    & &Privacy\\
    & &Safety guidelines for the employee\\
    & &Risk assessment\\
    \midrule

    \multirow{16}{*}{Governance} & \multirow{16}{*}{Data altruism and data sharing} & Transparency of the AI software\\
    & &Openness of the AI software\\ 
    & &Participation in sustainability strategy development\\ 
    & &Participation in sustainability strategy approval\\ 
    & &Participation in sustainability strategy implementation\\ 
    & &Leading role for sustainability and/or team\\
    & &Committee in charge of sustainability\\
    & &Sustainability reporting\\
    & &Ethics guidelines\\ 
    & &Participation in AI guidelines development\\ 
    & &Participation in AI guidelines approval\\ 
    & &Participation in AI guidelines implementation\\ 
    & &Reporting of AI usage\\
    & &Leading role for AI\\
    & &Committee in charge of AI\\
    & &Proactive compliance regulatory watch\\

    \midrule
    
    \multirow{9}{*}{Transvers} & \multirow{9}{*}{Global sustainability} & Sustainability definition, vision, model and understanding\\
    & &Sustainability in the company's culture\\
    & &Sustainability analysis\\ 
    & &Sustainability strategy and goals\\
    & &Sustainability transvers actions\\ 
    & &Sustainaility research and innovation\\ 
    & &Sustainability life cycle\\
    & &Positive-impact products (reduce carbone emission, ressource consumption, etc.)\\
    & &Donation and volunteer work for sustainability\\
    
    \bottomrule
  \end{tabular}
  }
\end{table*}

\subsection{Weights indicators and dimensions}

This section discusses differences in steps during the aggregation of indicators using weights and preparation of indicators prior to aggregation. 

\subsubsection{How we select indicators}

In its overarching structure, the evaluation model adopts the Doughnut framework, as introduced by K. Raworth~\cite{raworth2012safe}, encompassing an environmental ceiling, a governance, and a social floor. Together, these demarcate a conceptual doughnut's outer and inner boundaries, delineating a sustainability space for the conduct of the evaluated organization's activities. The categories governance and global sustainability are added to this framework to account for the third dimension of the ESG criteria (environmental, social, and governance) and the cross-cutting strategies and actions undertaken to transition their activities towards enhanced sustainability. Governance is based on the OECD AI Principles and the European Proposal on AI Act. It also refers to ISO norm 26000 on sustainability and ISO norm on AI ethical and societal concerns (ISO/IEC TR 24368:2022). 
The indicators grouped within the environmental ceiling draw inspiration from the study on planetary boundaries by Rockström et al.~\cite{rockstrom2009safe}, as adopted by Kate Raworth~\cite{raworth2012safe}. However, they are tailored to the specific context of assessing the sustainability of AI usage. In this regard, four indicators have been selected to represent the impact of AI usage best while ensuring a manageable level of complexity and detail in line with the information available to potential respondents. Each indicator aligns with a sustainability vision but possesses its unique rationale, as detailed below:
\begin{itemize}
    \item Climate Change: This indicator was selected based on the significance of energy consumption in evaluating the sustainability of digital applications, including AI, which subsequently results in direct and indirect greenhouse gas emissions.
    \item Natural Resources: This indicator was selected to reflect the impact of often overlooked infrastructures supporting digital technologies, especially AI. It should be noted that this indicator complements the planetary boundaries perspective, as the depletion of natural resources (considered more broadly than freshwater) does not jeopardize the overall equilibrium of the Earth's system. Instead, it only threatens the sustainability of access to specific strategic resources for our economy. Despite its omission in the planetary boundaries model, integrating non-renewable natural resources is expected in sustainability assessments and aligns with circular economy strategies.
    \item Pollution and Disruption of Natural Cycles: This indicator was chosen to encapsulate planetary boundaries related to introducing novel entities, disrupting biogeochemical cycles, and accumulating aerosols in the atmosphere. Within assessing AI sustainability, these three impact areas can be simplified under the pollution concept, which is better understood and quantified by organizations.
    \item Biodiversity and Ecosystems: This indicator directly referenced the "biosphere integrity" planetary boundary. It was chosen to reflect a potential organization's consideration of its impact on biotopes and the biocenosis forming their environment. It is worth noting the exclusion of factors such as "Atmospheric Ozone Depletion," "Land-System Change," and "Ocean Acidification." These were deemed less relevant for assessing the sustainability of AI usage and too detailed for a questionnaire.
\end{itemize}

The indicators grouped within the social floor incorporate the health and well-being, privacy, and education factors from the Doughnut model, deemed relevant for organizations. These are complemented by security, considered a core element of the social dimension of the ESG for organizations. Together, these three indicators provide a relevant framework for evaluating the human aspects of sustainability in the context of AI use.

\begin{figure}[!b]
  \centering
\includegraphics[width=1\linewidth]{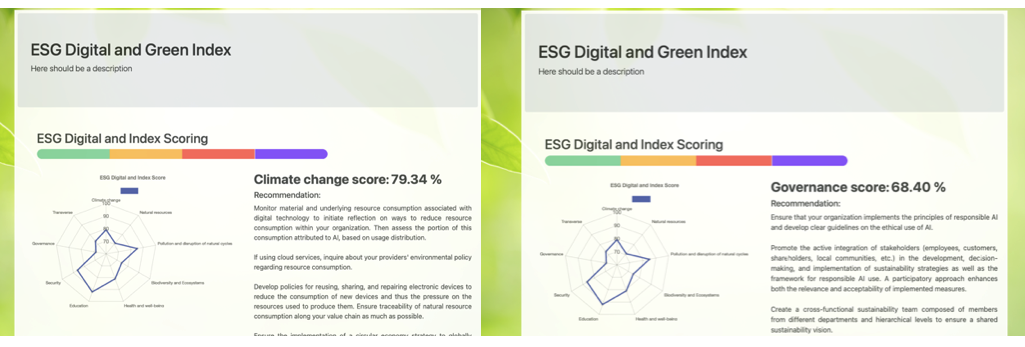}
  \caption{The ESG Digital and Green Index Dashboard.}
  \label{fig:dashboard}
\end{figure}

\subsubsection{Normalization}

Normalization is crucial when developing a composite index, mainly when the index builds on multidimensional concepts and covers many indicators. It helps to transform indicators with different units into uniform scales and unitless numbers that allow meaningful comparisons~\cite{pollesch2016normalization}; align indicators with positive and negative relationships to the phenomenon, which, in the case of this report, is green AI growth~\cite{mazziotta2013methods}; and reduce uneven influence of indicators with extreme values on the index~\cite{talukder2017developing}. The most common normalization methods include ranking, distance to target, or the best performer; standardization, or z-scores; re-scaling, or min-max transformation; and proportionate normalization. 

\subsubsection{Aggregation of indicators and dimensions}

For many green AI growth indicators, such as efficient and sustainable use of resources and natural resources and green economic opportunities and governance, mathematical models are available, represented by deductive models to explain the behavior of the systems. The described approach is based mainly on statistical analyses to identify the indicators' relationship to relevant explanatory variables. Aggregation reduces dimensionality and provides a holistic value to measure performance ~\cite{molnar2022modeling}. The two most common and straightforward methods include linear aggregation using the arithmetic mean and geometric aggregation using geometric mean~\cite{molnar2022modeling}, with the former being more widely applied than the latter. In our model, we use the weighted arithmetic mean method with the weights corresponding to indicator categories. Thanks to that, rational weights are given to each indicator category or item explicitly, and these weights indicate the relative importance of the indicators included in the determination of the index.

\subsection{Implementation}

We implemented a proof-of-concept system that calculates the DGI for each indicator category (\cref{fig:framework}). Our model was implemented as an application\footnote{The GitHub repository of the ESG Digital and Green Index is anonymized as it may reveal the identity and affiliation of the authors. We will provide the link upon acceptance.} using the Python programming language. \cref{fig:application} presents a screenshot of the application for the ESG Digital and Green Index with survey views of each indicator category. The reasons for choosing the Python program to build the application for the DGI include its compatibility with different platforms such as Windows, Mac, and Linux. Moreover, it is a readable and understandable language and has a language library. Python makes it easy to learn a language, thus reducing the time for production. It is known for its solid and high performance and free of charge, thus allowing a good community of practice sharing knowledge and libraries. Another important consideration for choosing Python is its versatile applications, including web and desktop apps, thus allowing the development of a graphical interface for showing simulation results with recommendations (\cref{fig:dashboard}), and complex calculation systems, facilitating the integration of different mathematical models for the interlinked systems of the green AI growth dimensions. 

After the application of the DGI, stakeholders will be trained to run scenarios on the application and use it for different planning and policy purposes. An interactive interface will be created based on the scenarios identified by the stakeholders.

\section{Discussion}
\label{sec:discussion}

In this paper, we present the ESG Digital and Green Index (DGI), a comprehensive framework that enables diverse stakeholders to assess the AI development, deployment, use, and maintenance pipeline across dimensions of sustainability. 
Our proposed framework, developed in a collective interdisciplinary effort involving researchers from law, sustainability, AI, and HCI, seeks to provide a holistic tool --deployed as an application-- to reveal the latent and nuanced aspects of AI use, not only on the environment and finite earthly resources but also our societal and personal well-being in addition to governance aspect of digital technologies. 
The following sections reflect on our framework and discuss the broader implications for the AI and HCI community.

\subsection{Addressing the Elephant in the Room -- The Case of AI Sustainability}

The DGI takes a multi-dimensional and multi-layered approach to consolidating the broader sustainability impacts of AI systems. 
Unlike existing metrics that capture the environmental impact of digital systems --including AI-- in terms of simplistic and rather incomplete indicators such as carbon footprints, the DGI extends this methodology by amalgamating broader environmental and societal impact as well as the aspects of governance of digital and AI systems.
We believe that such a comprehensive framework will enable organizations and the various stakeholders involved in the development, deployment, and use of AI systems to assess the sustainability impacts of their systems and, in turn, contribute to the development of processes and best practices that can facilitate the transition to a responsible and sustainable digital ecosystem. 
In this way, this article contributes to the fields of AI and sustainability by presenting a rich framework for quantifying an organization's impact in terms of energy, natural resources, societal and ethical impacts, and responsible governance of technological advancements related to its innovative AI-powered digital products.
More importantly, our work contributes to the HCI domain by \textit{a)} positioning its dissemination at the intersection of societal and technological realms and their collective evolution and \textit{b)} equipping HCI researchers with a sandbox to test and improve the broader sustainability impact of AI ecosystem, in a participatory manner with the diverse stakeholders.

Although presented as a \textit{generic} self-assessment tool, the DGI aims to raise awareness among different stakeholders about the problems associated with the training and use of massive models, such as Transformers and LLMs, in the short term.
In the medium and long term, however, we expect researchers to refine and extend the framework for different contexts, sectors, and stages of AI development and use. 
In addition, the DGI can enable policymakers, civil societies, and communities to develop and implement policies that promote sustainable practices using AI systems.
For instance, it can foster the development of a sustainability scoring systems for digital and AI systems, similar to the nutrition scoring system (e.g., Nutri-Score~\footnote{Nutri-Score: \url{https://en.wikipedia.org/wiki/Nutri-Score} (last visited on 13/09/2023).}) used for food items.

\subsection{Blending Societal and Ethical Aspects with Sustainability}

The conventional paradigm of sustainability and sustainability assessment has focused only on the environmental impact --manifested as a carbon footprint-- and the broader life cycle analysis of the components that make up these systems.
Through the development of the DGI, we have essentially diverged from this conventional perspective, and we have significantly expanded on the assessment of sustainability by bringing in facets of \textit{a)} impact on biodiversity and ``more-than-human'' ecosystems, \textit{b)} regulation and governance of the entire digital ecosystem, \textit{c)} education of stakeholders engaged in the digital ecosystem (i.e., all actors who are involved in the development, deployment, procurement, and use of such systems), \textit{d)} personal and societal well-being, and \textit{e)} alignment with the UN's SDGs. 
In this way, we aim to raise awareness among researchers, technology developers, policymakers, and end-users and to facilitate transparency about the visible, direct, nuanced, and peripheral effects of digital and AI systems.
By highlighting these cross-pillar dependencies and interactions in the sustainability assessment, the DGI enables the enrichment of ``responsible'' and ``ethical'' benchmarks for digital and AI development. 
In addition, it provides alternative discourses that define the future research agenda around the sustainable development and use of digital and AI systems while offering new ways to scrutinize existing human-centered methodologies and transform them into a planet-centered approach.

\subsection{Limitations}
Our study findings may be susceptible to limitations due to the methodology chosen, the novelty of the topic, and the pace of development of AI-based technologies. Firstly, regarding the methodology, the decision to insert a question or not can be seen as arbitrary despite having used a deliberative methodology via a group of experts in a multidisciplinary group. We decided to choose ISO norm 26000 and legal frameworks as the basis for the questions in addition to the Doughnut methodology. Interviews of experts representing different groups of users and interests may be relevant. It would allow us to identify meaningful use cases. Weights determine the relative importance of the indicators to each other. It entails the use of expert or subjective judgment that can become complicated in the case of a multidimensional concept.

The weighting of indicators can also be perceived as arbitrary based on personal judgment. Therefore, it would be meaningful to identify studies to justify this distribution key. The following lines clarify the issue: 
50\% Environmental ceiling, 20\% social floor, Governance 15\%, Global sustainability 15\%.  Environmental ceiling:  50\% climate change; 35\% natural resources; 10\% Pollution; 5\% biodiversity. Social floor: 33.3\% Health and Well-being; 33.3\% Education; 33.3\% Security.

An ongoing literature review is also central to gaining knowledge of the international legal framework and recent initiatives in this field. In particular, the International Sustainability Standards Board (ISSB) publication is a cornerstone. The initiative taken in the summer of 2023 by ISSB to develop educational material to explain and illustrate how an entity might apply some requirements to disclose information about some natural and social aspects of climate-related risks and opportunities is also very inspiring for our work.

\section{Conclusion}

In this paper, we discussed the rationale and value of a comprehensive framework to assist public and private organizations in evaluating the sustainability of AI software and digital ecosystems. This paper presented a historical perspective of AI and sustainability and existing frameworks for assessing AI and digital sustainability. It also identified gaps in current approaches.

This framework integrates environmental, societal, and economic considerations to foster responsible technological development and use and ensure a positive impact on society and the environment. Environmental sustainability addresses the challenges of energy efficiency and resource utilization (e.g., hardware efficiency, data center footprint), carbon footprint and emissions (e.g., greenhouse gas emissions, life cycle analysis), as well as the impact on the ecosystem (e.g., biodiversity). Social sustainability focuses on ethical considerations and responsible AI principles (e.g., fairness and bias mitigation, privacy and data protection, explainability, transparency, etc.), on inclusivity and accessibility (e.g., accessibility for diverse populations, digital divide mitigation), and human rights (e.g., social responsibility in AI development, impact on workforce). Finally, it also addressed economic sustainability in questioning an AI-based economy's economic viability and longevity. It focuses on the ethics of business model and business model resilience, cost-effectiveness, and return on investment. As we can see with LLM, innovation and technological advancements require continuous improvement to gain efficiency and reduce costs. Future-proofing strategies will require a cost-benefit approach to remain sustainable and socially accepted. This paper shows that the reference to Doughnut's model was a rational choice to consider the interest of all stakeholders, particularly the limited natural resources available. It is a call to action for industry stakeholders, policymakers, and researchers to invest resources in this new area of research: computational sustainability.

As our research shows cross-pillar dependencies and interactions, studying use cases will be the next step to refine the tool. 
Due to the proliferation of generative AI, which offers both opportunities and challenges for HCI, practical applications in this field will be our next focus to study further the intricate relationship between generative AI, HCI, and finite resources on Earth, shedding light on their interplay and implications. LLMs are transforming Human-Computer Interaction, making technology more intuitive and user-friendly. However, addressing the resource allocation dilemmas and ethical considerations associated with Generative AI is crucial to ensure a sustainable and responsible use of future technology. This paper contributed to raising the awareness of the HCI Community towards the urgent need to reflect on the impact of Generative AI models scaling on HCI. The scaling of Generative AI models will continue until morale improves.

\bibliographystyle{plain}
\bibliography{arxiv}  %

\end{document}